# Single cell data explosion: Deep learning to the rescue

The plethora of single-cell multi-omics data is getting treatment with deep learning, a revolutionary method in artificial intelligence, which has been increasingly expanding its reign over the bioscience frontiers.

A.K.M. Azad & Fatemeh Vafaee*

The traditional 'bulk' molecular profiling *averages* out information across different cell types within a population of millions of potentially heterogenous cells[1]. This composite measure masks individual variations among cells and limit our understanding on cell-specific functions[2]. In contrast, like a new magnifying lens, single-cell technology has enabled researchers to review biological hypotheses with high-resolution insights zoomed at the single cell level, offering monumental opportunities in functional biology, precise diagnosis and precision therapy[2]. Single-cell data analysis can reconstruct continuous cell lineages, identify rare cell subtypes and predict novel cell types in a heterogeneous tissue[3]. It enables an understanding of an individual cell's functions in the context of its micro-environment[4], unmask variations between individual disease-associated cells[2], and facilitates the discovery of molecular signatures accounting for different cellular states[1]. Nonetheless, the enormously growing volume of single-cell data comes with several challenges, i.e., curse of dimensionality, sparsity, degree of noise, batch errors, and data heterogeneity[5] which often hinder the performance of conventional computational approaches to scale up as data complexity and size grow, setting the stage for the contemporary *deep learning* algorithms.

Deep learning, a powerful subfield of machine learning, is now captivating the recent hype in the artificial-intelligence industry through its promises in near-human-level judgments. Following its remarkable achievements in the field of image recognition and natural language processing, deep learning is now effectively drawing the attentions of biotechnology researchers with a broad range of successful applications spanning *from* the biomarker discovery in medical images[6] *to* predicting functions of non-coding mutations in DNA sequences[7].

Architecturally, deep learning models have multi-layer *artificial neural networks* at their core, initially inspired by the structure and functions of the brain. In deep-learning networks, each layer consists of several nodes (i.e., computational units) coupled with a non-linear 'activation' function transforming large volume of input data into increasingly more abstract feature representations. The model is eventually trained by iterative parameter readjustments by relating those features with the known labels in order to make predictions on unseen data. Although training deep learning models possess computationally intensive operations for their numerous learnable parameters, Graphics processing units (GPUs) can be used to parallelize its major operations, i.e. the matrix multiplications required at their training phase[8].

The power of deep neural networks is in *automatic feature learning* from *massive* datasets which not only streamlines laborious manual feature selection but also facilitates learning task-optimal features[8]. They have achieved far better performance accuracies compared to classical machine learning methods in many complex data domains, with their performance scaling up effectively as data size grows. Overall, methods using deep-learning architectures can effectively deal with *large-scale*, *heterogeneous* and *complex information* which is specially demanded in the analysis of single-cell sequencing datasets[9].

Single-cell sequencing technologies can now simultaneously assay hundreds of thousands of individual cells provoking a "big data" phenomenon[10]. Moreover, for a more comprehensive understanding of cellular phenotypes, sequencing technologies are now offering simultaneous profiling of genome, epigenome, transcriptome, and proteome of an individual cell igniting further data explosion as single-cell *multi-omics*[11]. Processing and interpreting such high-dimensional single-cell information increasingly challenges conventional computational informatics[12] calling for powerful and scalable deep learning models[13] for dropout imputation, cell-subtype clustering, phenotype classification, visualization, and multi-omics integration (Figure 1).

Exploring the "big" single cell data requires the reduction of its input dimensionality in order to visualize data in low-dimensional space in a scalable manner. A deep learning-based method, called Sparse Autoencoder for Unsupervised Clustering, Imputation, and Embedding (SAUCIE) has been used to reduce input dimension of single-cell data by successive applications of narrower layers[5]. The SAUCIE architecture also incorporates special layers for visualizing data in two-dimensional space, removing batch-effects, and automated data clustering[5].

Despite the data massiveness, single-cell data is often zero-inflated implying that more than 70% of read counts are frequently observed as either truly zeros or "dropouts" (false negatives)—typically caused by the technical failure to detect minute amounts of DNA/RNA present in a single cell[14]. This reduces data quality, imposes bias, and impedes effective downstream bioinformatic analyses[14]. More specifically, the dropout events may also result in near-zero values for low-abundance genes, jeopardizing their confidence-levels for reliable statistical inference[15]. Several state-of-the-art imputation methods for single-cell data exists till date but major concerns remain with their long run-time and memory consumptions in the context of single-cell big-data. *DeepImpute*, a deep learning-based approach claimed noticeable enhancements in these regards by employing a set of deep neural sub-networks that lowers model complexity and over-fitting, with improved parallelism and enhanced accuracy among the competing methods[14].

Latterly, parallel profiling of multiple molecular types (e.g. DNA, chromatin, RNA and protein) of an individual cell has enabled an inclusive inference about cellular phenotypes and their associations with various diseases[11]. Integration of disparate multi-omics data poses various computational challenges, when properly addressed will impact medicine by enhancing health management as well as disease diagnosis and treatment[16]. Deep learning is an ideal model for multi-omics data integration[17] with multiple success stories in integrating bulk profiles for prognosis prediction and precision medicine[18,19], nourishing its prospect for integrating *single-cell* multi-omics data[11].

What makes deep learning usage more enriched is the availability of open-source libraries, such as TensorFlow, Theano, Caffe, PyTorch or MXNet with varying degree of ease-of-use in defining and compiling deep neural network models and training or testing models on dataset. However, a major concern around deep learning methods is the 'black-box' nature of the models and their un-interpretability due to the huge number of parameters and the complex approach for extracting and combining features. While data science community is active in enhancing interpretability of deep neural networks, further research in biomedical contexts are required to understand clinically or biologically relevant patterns in data raised to accurate predictions, and to improve the users' trust ensuring that the model decides based on reliable reasons rather than artifacts in data[20].

Nevertheless, deep learning is a promisingly potent machine learning technology, and the ongoing research in this field are expected to reign over the recent "big bang" of single-cell multi-omics data, just like it has been doing in other fields.

**A.K.M. Azad** *and* **Fatemeh Vafaee** *are in the School of Biotechnology and Biomolecular Science, University of New South Wales, Sydney NSW 2052, AU*
*Correspondence to f.vafaee@unsw.edu.au

1. Stegle, O. *et al. Nat. Rev. Genet.* **16(3)**, 133-145 (2015)
2. Shalek, A. K. & Benson, M. *Sci. Transl. Med.* **9(408)** (2017)
3. Deng, Y. *et al.* Preprint at *bioRxiv* https://doi.org/10.1101/315556 (2018)
4. Eberwine, J. *et al. Nat. Methods* **11**, 25-27 (2014)
5. Amodio, M. *et al.* Preprint at *bioRxiv* https://doi.org/10.1101/237065 (2017)
6. Wainberg, M. *et al. Nat. Biotechnol.* **36**, 829–838 (2018)
7. Zhou, J. *et al. Nat. Methods* **12**, 931–934 (2015)
8. Angermueller, C. *et al. Mol. Syst. Biol.* **12(7)**, 878 (2016)
9. Cho, H. *et al. Cell Systems* **7(2)**, 185 - 191.e4 (2018)
10. Iacono, G. *et al. Genome Res.* **28**, 878–890 (2018)
11. Macaulay, I. C. *et al. Trends Genet.* **33(2)**, 155–168 (2017)
12. Yu, P. & Lin, W. *Gen. Prot. Biol.* **14(1)**, 21-30 (2017)
13. Rusk, N. *Nat. Methods* **13**, 35 (2016)
14. Arisdakessian, C. *et al.* Preprint at *bioRxiv* https://doi.org/10.1101/353607 (2018)
15. Lun, A.T.L. *et al. F1000 Research* **5**, 2122 (2016)
16. Karczewski K.J. *et al. Nat. Rev. Genet.* **19**, 299-310 (2018)
17. Grapov D. *et al. OMICS: A Jour. of Int. Biol.* **22(10)**, 630-636 (2018)
18. Chaudhary, K. *et al. Clin. Canc. Res.* **24(6)**, 1248–1259 (2018)
19. Zhang, L. *et al. Fron. in Genet.* **9**, 477 (2018)
20. Ching, T. *et al. J. R. Soc. Interface* **15(141)** (2018)

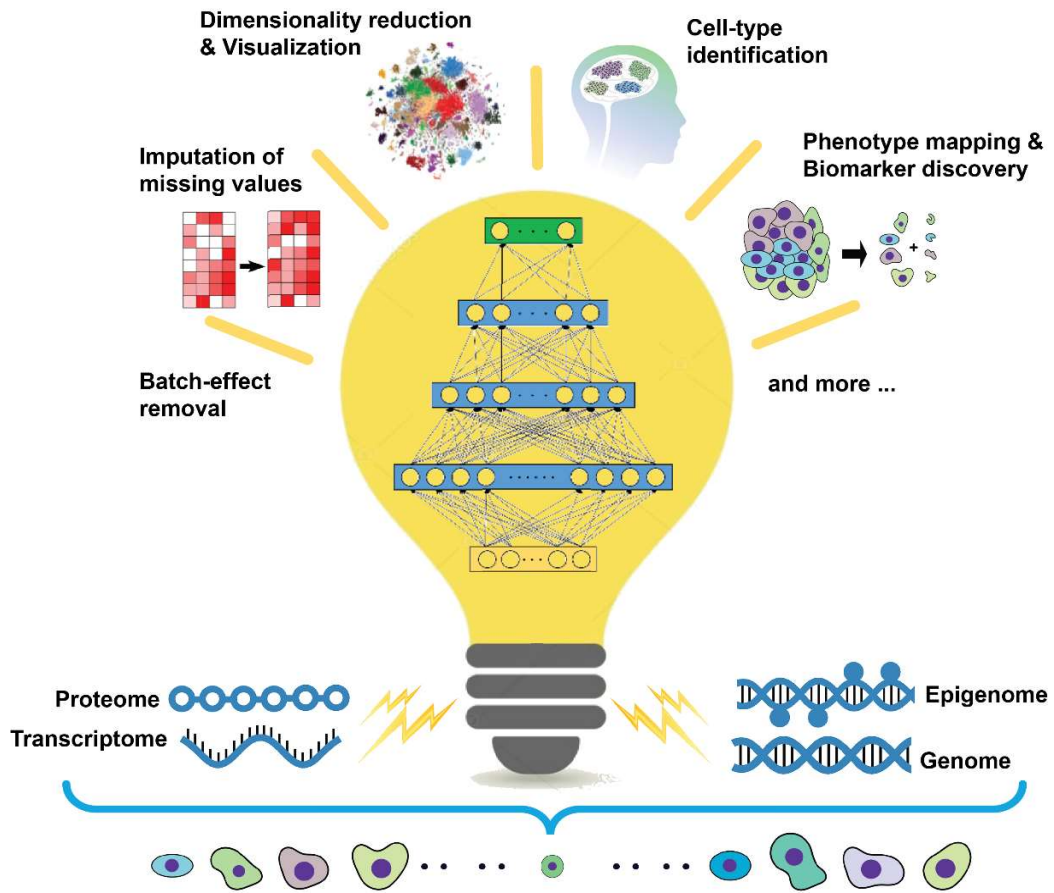

Figure 1: Deep learning is capable of offering multiple biological/clinical insights from single-cell multi-omics data